\documentclass[prb,twocolumn, superscriptaddress]{revtex4-1}
\usepackage{graphicx,amsmath,amssymb,amsxtra,epstopdf,wasysym}
\usepackage{hyperref}   
\renewcommand{\narrowtext}{\begin{multicols}{2} \global\columnwidth20.5pc}

\def\be{\begin{eqnarray}}
\def\ee{\end{eqnarray}}

\newcommand{\Eq}[1]{Eq.~(\ref{#1})}

\newcommand{\cC}{{\cal C}}

\newcommand{\ket}[1]{{|#1\rangle}}
\newcommand{\bra}[1]{{\langle#1|}}

\newcommand{\ZZ}{{$\mathbb Z_2$} } 
\def\RVB19{3280}
\begin{document}

\title{
Ground state uniqueness of the twelve site RVB spin-liquid  parent Hamiltonian on the kagome lattice
}

\author{Zhenyu Zhou}
\affiliation{
Department of Physics and Astronomy,
University of Pittsburgh, Pittsburgh, PA 15260, USA}
\affiliation{
School of Physics, Astronomy and Computational Sciences,
George Mason University, Fairfax, VA 22030, USA}

\author{Julia Wildeboer}
\affiliation{
National High Magnetic Field Laboratory, Florida State University, 
Tallahassee, FL 32310, USA}

\author{Alexander Seidel}
\affiliation{
Department of Physics, 
Washington University, St. Louis, MO 63136, USA}
\date{\today}

\begin{abstract}
Anderson's idea of a (short-ranged) resonating valence bond (RVB) spin liquid has been the first ever proposal of what we now call a topologically ordered phase.
Since then, a wealth of exactly solvable lattice models have been constructed that have topologically ordered ground states. For a long time,
however, it has been difficult to realize Anderson's original vision  in such solvable models,
according to which the ground state has an unbroken SU(2) spin rotational symmetry
and is dominated by fluctuation of singlet bonds.
The kagome lattice is the simplest lattice geometry for which parent Hamiltonians stabilizing a
prototypical spin-1/2  short-ranged RVB wave function has been constructed and strong evidence has been given that this state belongs to a topological
phase. The uniqueness of the desired RVB-type ground states has, however, not been rigorously proven for the simplest possible such Hamiltonian, which
acts on 12 spins at a time.
Rather, this uniqueness has been demonstrated for a longer ranged (19-site) variant of this Hamiltonian by Schuch et al., via making contact with powerful results
for projected entangled-pair states. In this paper, we extend this result to the 12-site Hamiltonian.
Our result is based on numerical studies on finite clusters, for which we demonstrate a ``ground state intersection property''
with implications for arbitrary system size.
We also review the relations between various constructions schemes for RVB parent-Hamiltonians found in the literature.
\end{abstract}

\maketitle
\section{Introduction} 

The study of topological phases of matter has become
a dominant theme of contemporary research into condensed matter physics.
The notion of a ``topological phase'' has become attached to a
wide range of systems, from non-interacting ones such as
Chern-\cite{haldane} and topological\cite{kanemele} insulators
to interacting systems displaying non-trivial topological orders.\cite{wenniu}
The latter have found a multitude of realizations in the physics of the 
fractional quantum Hall effect,\cite{tsui,laughlin} and are considered
possible hardware for topological quantum computation.\cite{kitaev1, freedman3}
However, the earliest incarnation of topological order and, more generally,
systems with fractionalized excitations that surfaced in the condensed matter physics literature
was proposed by Anderson in the context of quantum magnetism.\cite{anderson1,anderson2}
The key question asked in this seminal work was whether a quantum antiferromagnet
could retain all symmetries of the system in its ground state. These symmetries
were understood to include {\em both} the space group of the underlying lattice structure
{\em and} SU(2) rotational invariance in spin space, as well as time-reversal
symmetry. For this Anderson developed the scenario of the
so-called ``resonating-valence-bond'' (RVB) spin liquid.\cite{anderson1, anderson2}
While the term is used broadly for physically distinct situations, namely both long-ranged
or critical RVB spin liquids and short-ranged ones, it became clear through a 
number of subsequent developments
that the latter
are characterized by a number of interesting topological properties.
These include, in particular, semionic fractional statistics and non-trivial
topological degeneracy,\cite{kivelson, readchakraborty} 
which are nowadays easily recognized as the hallmarks of topological order.\cite{wenniu}
The topological quantum numbers that the short-ranged RVB spin liquid was argued to
have are those of the $\mathbb Z_2$ topological phase.

In the original RVB scenario, the spin liquid wave function is a superposition
of many different valence bond states, where each valence bond corresponds to a
singlet pairing between two spin $1/2$ degrees of freedoms localized on different lattice sites.
It proved difficult, however, to realized such a scenario in both realistic and toy-model Hamiltonians
for frustrated quantum anti-ferromagnets. Instead, initial successful demonstrations 
of the \ZZ topological phase was given in models that completely abandoned the RVB-constraint of
SU(2) invariance. 
The first such demonstration was Kitaev's the toric code model.\cite{kitaev1}
More true to the original RVB idea was the construction of quantum dimer models (QDMs),\cite{RK}
where the first demonstration of the \ZZ phase was given for the triangular lattice QDM
by Moessner and Sondhi.\cite{MS}
In QDMs, the valence bonds of RVB physics are mimicked by hardcore bosonic degrees of freedom
(``dimers''). However, there is no non-trivial way in which a global SU(2) symmetry is realized.
The recipe of Ref. \onlinecite{MS} was successfully generalized by several other works,\cite{misguich, balents}
but at first without reinstating SU(2) invariance. It remains non-trivial to address the general  question
where, if anywhere, within the phase diagram of SU(2)-invariant local spin 1/2 Hamiltonians 
a topological spin liquid can be stabilized. Intuitively, one might argue that
the breaking of symmetries becomes  the harder to avoid the more symmetries there are that could possibly be broken.
Positive results, however, were obtained early on for large-$N$ generalizetions of the SU(2) symmetry.\cite{readsachdev} 
Moreover, for SU(2) spins on a highly decorated lattice, a controlled procedure
was given in Ref. \onlinecite{RMS} to show that wave functions of the nearest-neighbor RVB
form can be stabilized by a local parent Hamiltonian.
By this we mean wave functions of the following kind:
\begin{equation}\label{RVB}
 \ket{\sf RVB}=\sum_D \ket{D}\,,
\end{equation}
where the sum goes over all dimerizations of the lattice into nearest neighbor pairs,
and $\ket{D}$ denotes a state where each pair of the dimerization carries a singlet,  following some sign
convention.
This settles the existence of general lattice structures that support topological spin liquids, for some choice of Hamiltonian.
However, one may still ask the same question with a given and fairly simple lattice in mind, such as the square, triangular, 
or kagome. It is then natural to seek
parent Hamiltonians for states of the form \eqref{RVB} defined for the specific lattice structure in question.
On general grounds, however, it is expected that \Eq{RVB}, with only nearest neighbor valence bonds appearing,
 describes a stable phase only 
on a non-bipartite lattice. 

On bi-partite lattices, any parent Hamiltonian for \Eq{RVB} will 
generally inherit the extensive ground state degeneracy of QDMs defined on bipartite
lattices.\cite{MSFradkin}. This extensive ground state degeneracy is intimately related
to the critical behavior of correlation functions.
This is demonstrated by the cases considered by Fujimoto,\cite{fujimoto}
where parent Hamiltonians for \Eq{RVB} for both the square lattice and the honeycomb
lattice were constructed (although the square lattice Hamiltonian of Ref. \onlinecite{fujimoto}
was deficient, in that it admitted exponentially more ground states than intended, as
pointed out by Cano and Fendley,\cite{cano} who gave a valid construction for the square lattice and a simplified one for the honeycomb).
Indeed, the local parent Hamiltonians known to stabilize \Eq{RVB} for these lattices  have (at least)
the full extensive (multi-critical) ground state degeneracy of the associated QDMs.
 Moreover, recent Monte-Carlo studies\cite{alet,sandvik} of correlation functions in the singlet sector of \Eq{RVB} on the square lattice
 demonstrate the critical behavior that is predicted qualitatively by the QDMs on the same
 lattice. In contrast, for non-bipartite lattices,
 QDMs have exponentially decaying correlations, and, over some range of parameters,
 no broken symmetry combined with the topological degeneracy expected of the
 \ZZ phase.\cite{MS, misguich}
 
 \subsection{RVB parent Hamiltonian(s) on the kagome lattice}

The above considerations motivate the search for parent Hamiltonians for the $\ket{\sf RVB}$ state
on simple non-bipartite lattices. 
To the best of our knowledge, the first such has been given by one of us in Ref. \onlinecite{seidel}.
 Subsequent studies have  strongly supported the exponentially decaying nature of correlation functions\cite{schuch12, wildeboer, yao}
 and the liquid, symmetry unbroken character of the ground state.\cite{wildeboer}

In addition to showing these properties for the ``special'' ground states at the solvable point, one desires evidence that said properties are
stable to perturbation. Ideally, one would prove that the parent Hamiltonian has an energy gap,
as befits a topological  phase. This, however, is usually difficult to achieve in two and higher dimensions, unless the
Hamiltonian is so tuned that it is the sum of commuting operators.
On the other hand, the exponential decay of correlation functions, together with the fact that the Hamiltonian 
has the correct (finite) ground state degeneracy for any finite lattice size, is widely regarded strong circumstantial
evidence for the presence of an energy gap.
The uniqueness of the desired ground states, modulo  topological degeneracy, is also an essential ingredient  
in making the case that the parent Hamiltonian is not ``sick'' in the sense that it admits many more ground states than intended, perhaps extensively so or worse.
In Ref. \onlinecite{seidel}, it was argued that the most likely subspace to harbor additional ground states
is the space spanned by the nearest neighbor valence bond state $\ket{D}$.
It was shown there that this subspace does not contain any ground states besides
those of the form \eqref{RVB}, where, in the presence of periodic boundary conditions, the sum may be restricted to valence bond
states within one of four topological sectors.\cite{readchakraborty}
This result made use of the linear independence of the set $\ket{D}$, the general question of which is a long-standing problem
in the field of short-range RVB physics. In Ref. \onlinecite{seidel}, this question was answered positively for the kagome,
and for many other lattices in Ref. \onlinecite{wildeboer_li}, expanding earlier results.\cite{kcc}

\begin{figure}[t]
\centering
\includegraphics[width=8cm]{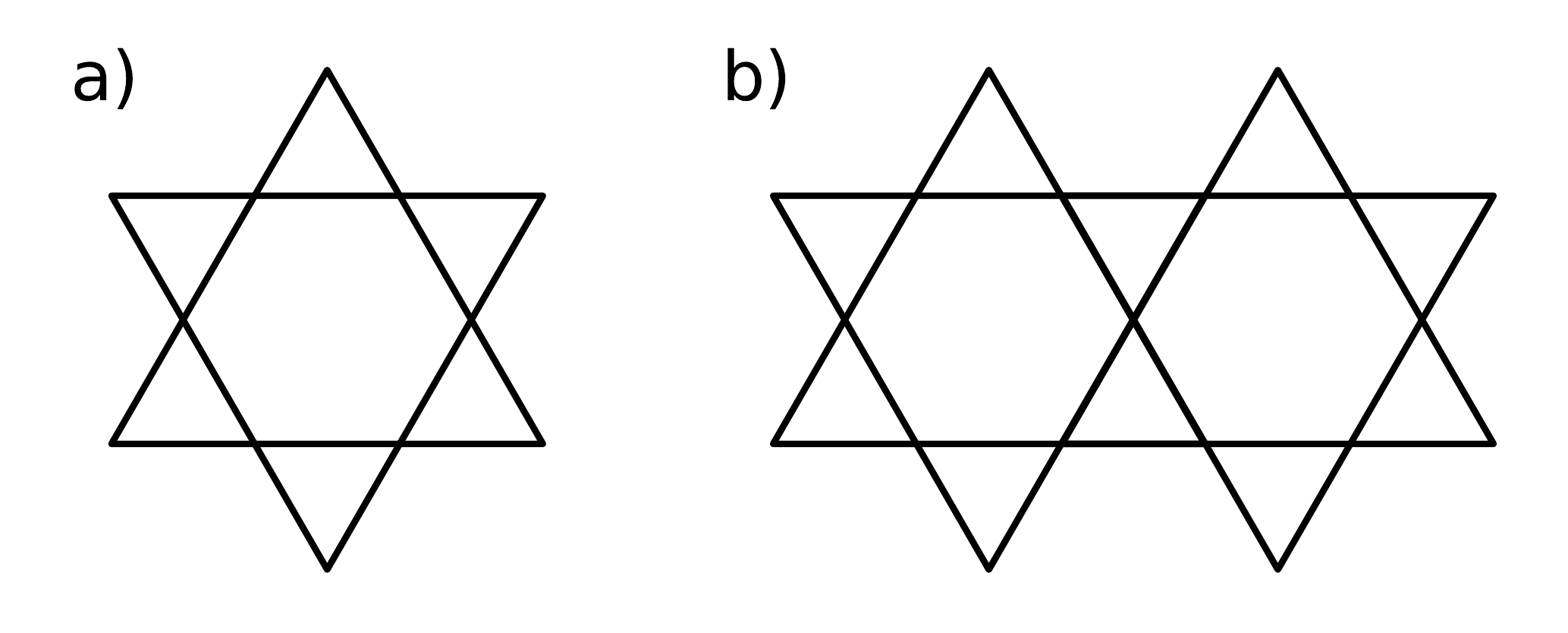}
\caption{
\label{cells}
 The 12-site ``star'' cell, a),  and the 19-site ``double star'' cell, b). 
 The kagome lattices we consider can be covered by either type of cell,
 and various parent Hamiltonians for the RVB state \Eq{RVB}
 are discussed that are given by sums of local terms acting 
 simultaneously on spins within cells of either type a) or b). 
}
\end{figure}

The result for the ground state uniqueness as given in Ref. \onlinecite{seidel}
is partial, in that it applies strictly only to the subspace spanned by
 nearest neighbor valence bond states. It does, however, apply to
a large class of parent Hamiltonians defined in the same paper,
which can be labeled by the basic local cell $\cC$ on which local
terms in the respective parent Hamiltonian act. The smallest possible
cell $\cC$ for which the construction of Ref. \onlinecite{seidel} yields
a non-trivial parent Hamiltonian is the twelve-site cell shown
in Fig. \ref{cells}a), and we will refer to the corresponding parent Hamiltonian 
as $H_{12}$. The next larger cell to which this construction can be meaningfully applied is
the 19-site cell of Fig. \ref{cells}b), and we will refer to the corresponding parent Hamiltonian
as $H_{19}$. The construction, which we review in Sec. \ref{hamiltonians},
makes it obvious that the larger the basic cell, the more likely the ground state uniqueness
will hold within the full Hilbert space. As of now, a rigorous proof of this uniqueness exists
already
for $H_{19}$.\cite{schuch12} The key insight leading to this result was the realization that
the RVB wave function \eqref{RVB} can be understood as a tensor network or projected entangled pair state (PEPS),
and builds on powerful criteria by Schuch et al.\cite{schuch10} for the ground state uniqueness of general
 PEPS parent Hamiltonians.

For $H_{19}$ (though not for some larger basic cells also considered),
the result of Ref. \onlinecite{schuch12} also makes use of the linear independence
property proven in Ref. \onlinecite{seidel}.
 Here, in turn, we will show that the ground state uniqueness, modulo topological degeneracy, of $H_{19}$ also
 implies that of  $H_{12}$. The latter had been conjectured earlier
 by one of us.\cite{seidel}
 This then completes the demonstration of the catalog of desirable
 properties discussed above for the ``simpler'' kagome lattice
 RVB parent Hamiltonian $H_{12}$.
Our results should also be of interest in the general context of the study
of frustration free lattice Hamiltonians and their ground state spaces
that has been of much interest recently.\cite{bravyi, schuch10, beaudrap, yoshida, michalakis,chen12}

The paper is organized as follows.
In Section \ref{hamiltonians}
we review the constructions of parent Hamiltonians
for the kagome lattice RVB state as done in Ref. \onlinecite{seidel}
using local RVB configurations
and in Ref. \onlinecite{schuch12} using properties of PEPS states,
and show that they are identical.
In section \ref{results} we discuss numerical studies on small clusters,
which, using properties of frustration
free Hamiltonians, imply  that for kagome lattices of any size, the ground state
spaces of the two Hamiltonians $H_{12}$ an $H_{19}$ are the same.
We conclude in Section \ref{conclusion}.

\section{\label{hamiltonians} Various constructions schemes for RVB parent Hamiltonians}

\subsection{Defining local RVB states and parent Hamiltonians\label{locRVB}}

A standard procedure to construct a pair  $(\ket{\psi}, H)$ consisting of a ground state $\ket{\psi}$ 
describing, by assumption, a certain phase and a parent Hamiltonian $H$
is to attempt to make this pair ``frustration free''.
This means that $H=\sum_i h_i$ is the sum of not necessarily commuting local terms $h_i$, such that
$\ket{\psi}$ is a common ground of each $h_i$. It is then necessarily also a ground state of $H$.
To determine the degeneracy of this ground state is usually a non-trivial problem that will be of interest
in the following. This question only depends on the ground state spaces of the individual $h_i$, which therefore may 
be taken
 to be the negatives of the projection operators onto their respective ground state spaces, or $h_i=-P_i$.
 
\begin{figure}[t]
\centering
\includegraphics[width=8.5cm]{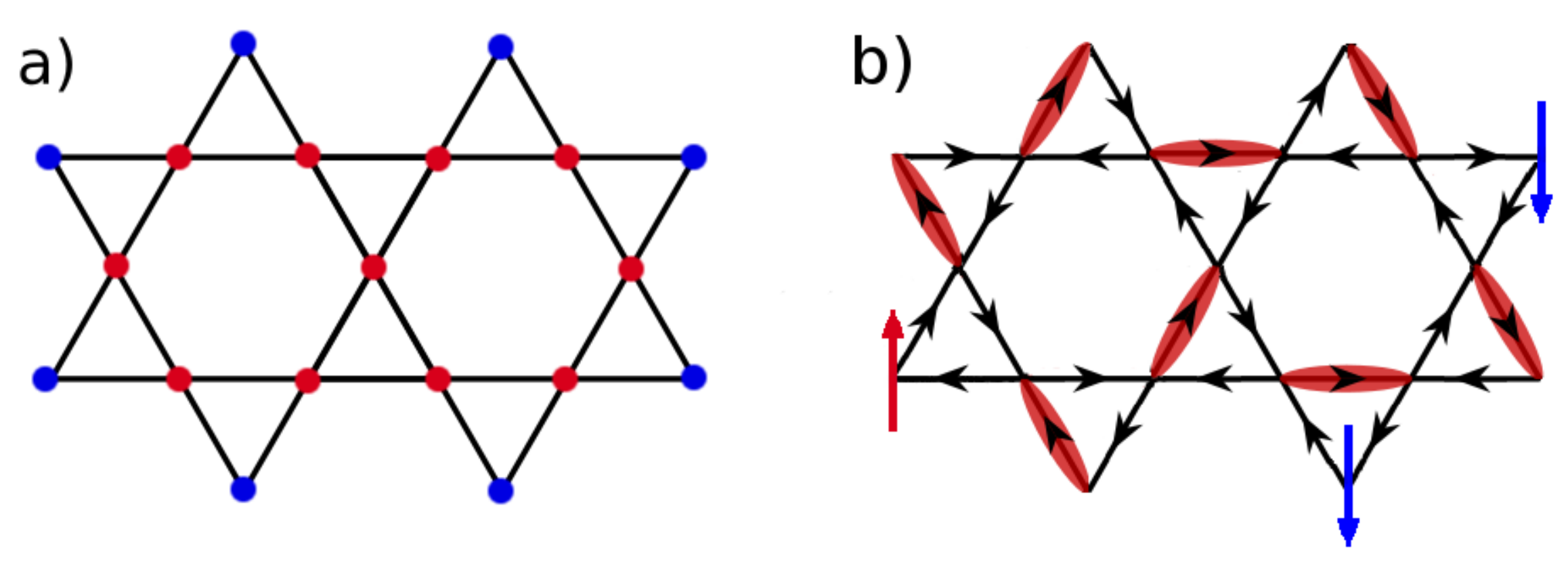}
\caption{
\label{pic2}
[Color online] 
a) inner (red) and boundary (blue) sites of the 19-site cell. 
b) A local valence bond state. All inner sites are required
to participate in valence bonds (ovals). Boundary sites may or may
not do so. Boundary sites not participating (free boundary sites)
are put in either an up- or a down-spin state.
The set of all local valence bond states satisfying the conditions
defined above and in the text are linearly independent for the
19-site cell,\cite{seidel} and come in dynamical equivalence classes (see text).
Equal amplitude superpositions within one class generate the subspace
of local RVB states, \Eq{psiDf}, which underlies the definition of the
parent Hamiltonian $H_{19}$ (see text). The sign convention used
to resolve the relative phases of different singlet configurations is indicated by arrows.
Analogous constructions are possible for various other cells, in particular
the 12-site sell, and other sign conventions, where the linear independence may or may not hold.
}
\end{figure}

One generally desires the ground state degeneracy (of $H$) to be small, especially when describing a stable phase.
For this the rank of the operator $P_i$, viewed as acting on the local Hilbert space of a small cell, should not be
too large. In contrast, when choosing all $P_i$ to be the identity operator, every state $\ket{\psi}$ would be a ground state,
and such a Hamiltonian would of course be entirely trivial. Somewhat in between these two extremes
are the so-called Klein models,\cite{Klein} whose ground states are exponentially degenerate in the system size,
and include the subspace of nearest neighbor valence bond states $\ket{D}$. Here one is interested 
on conditions such that the ground state space is exactly spanned by the exponentially many states $\ket{D}$.
This is the case for some lattices,
\cite{kcc} but not for others. Generally, it is thus desirable to have the rank of $P_i$ 
as small as possible. The condition that the state $\ket{\psi}$ is a ground state of $h_i$
is equivalent to the statement that the support of the local density matrix $\rho_i$ is contained in the
local ground state subspace of the operator $h_i$ (whose dimension is the rank of $P_i$). Here, by $\rho_i$ we mean the
density matrix  obtained by tracing $|\psi\rangle\langle\psi|$ 
over the complement of the local cell on which $h_i$ acts, and by the support of $\rho_i$, we mean
the orthogonal complement of its kernel, that is, the direct sum of its non-zero eigenvalue eigenspaces.
It follows that the smaller the rank of $P_i$, the more restrictive is the condition for a state to be the simultaneous
eigenstate of all the operators $h_i$. For given state $\ket{\psi}$, the smallest possible rank that $P_i$
can have is obviously equal to the dimension of the support of $\rho_i$, corresponding to the case where
$P_i$ is simply the projection operator onto said support.
Hence, to construct a parent Hamiltonian for given $\ket{\psi}$ in this manner, it is necessary to identify local
cells for which the local density matrix $\rho_i$ derived from $\ket{\psi}$ does not have full rank.
The smaller the rank of $\rho_i$ (the dimension of its support), the more likely it is that the state
$\ket{\psi}$ is the unique ground state of the parent Hamiltonian thus constructed, or at least is one of a fixed number of degenerate ground states,
independent of system size.


In Ref. \onlinecite{seidel}, a general recipe was given how to construct a local subspace
${\sf RL}({\cal C})$ of ``resonant loop'' states that necessarily contains the support of the local density matrix
$\rho_{\cal C}$ for the RVB state, $\ket{\sf RVB}$, for a given lattice with local cell $\cal C$.
Here we briefly review this recipe. Consider a local cell $\cC$, as depicted in Fig. \ref{pic2}b),
which is part of some larger lattice. For such a cell, we define a local dimerization as a pairing
of sites of the cell into nearest neighbors, such that each inner site participates in a pair (dimer),
but boundary sites may or may not. Here, by inner sites we mean sites for which all nearest neighbors
in the lattice topology also belong to the cell, whereas boundary sites of the cell do not satisfy this property.
An example for a local dimerization is also shown in Fig. \ref{pic2}b).

We will refer to the boundary sites not participating in a local dimerization as the
``free sites'' of that dimerization.


We may regard two local dimerizations $D$, $D'$ as ``dynamically equivalent'' if one
can be transformed into the other by certain local dimer moves. These dimer moves
are the same as those appearing in the Hamiltonian of a suitable quantum dimer model
on the same lattice. For the kagome lattice, they consist of moves shifting dimers long closed paths
that contain exactly one hexagon.\cite{misguich} These dimer moves have the benefit of
being ergodic within topological sectors. Note that for local cells and local dimerizations
as defined here, dimer moves never change the set of free sites.
We may therefore define local ``resonant loop'' state as a state of the form
\begin{equation}\label{psiDf}
  \psi_{[D],f}= \sum_{D' \in [D]} \ket{D'}\otimes \ket{f}
\end{equation}
Here, by $[D]$ we mean a dynamical equivalence class of local dimer coverings
of the cell $\cC$, and $\ket{f}$ denotes a {\em fixed} spin configuration of the
free boundary sites of $[D]$. The ket $\ket{D'}$ denotes a state where each pair in 
$D'$ carries a singlet, and the overall sign of the resulting product of singlets 
is determined by the link orientation also shown in Fig. \ref{pic2}b).
The desired subspace ${\sf RL}({\cal C})$ is then the linear hull 
of all possible states $\psi_{[D],f}$,
\begin{equation}
  {\sf RL}({\cal C})= \langle \{ \psi_{[D],f}   \} \rangle\,,
\end{equation}
for all possible dynamical equivalence classes $[D]$ {\em} and,
for given $[D]$, all possible free site configurations $f$ (a complete set of $\ket{f}$s).
The local subspace  ${\sf RL}({\cal C})$ contains the support of the 
local density matrix of the state $|{\sf RVB}\rangle$ for the cell $\cC$.\cite{seidel}
This is also true in the presence of periodic boundary conditions,
where one has four degenerate ground states, one for each 
of four topological sectors $[i=1\dots 4]$:\cite{readchakraborty}
\begin{equation}\label{RVBi}
 \ket{\sf RVB,i}=\sum_{D\in [i]} \ket{D}\,,
\end{equation}
where presently, $D$ again denotes a global dimerization of the lattice. 
We introduce the orthogonal projection operator $P_\cC$ onto the space ${\sf RL}({\cal C})$.
We are interested in conditions where the states
$\ket{{\sf RVB},i}$ are unique ground states of the parent Hamiltonian
\begin{equation}
   H=-\sum_{\cC} P_\cC\,, 
\end{equation}
where the sum goes over all cells of a certain ``type'' of geometry.
The smallest type of cell for which one may hope that this is the case
is the 12-site star shaped cell of Fig. \ref{cells}a). We will call the corresponding
parent Hamiltonian $H_{12}$. The purpose of this paper is to establish that
the unique ground states of $H_{12}$ are the four-fold degenerate states
\eqref{RVBi}. We note that in this case, as the cell contains only a single hexagon,
every local dimerization $D$ is dynamically equivalent to a single partner $D^\ast$.
We should therefore think of the space ${\sf RL}({\cal C}_{12})$ as
\begin{equation}\label{RVB12}
   \langle \{ (\ket{D} + \ket{D^\ast} ) \otimes \ket{f} \} \rangle\,.
\end{equation}
The other case of special importance is that of $H_{19}$, where the geometry
of the underlying cell is that of the 19-site ``double star'' cell of Fig. \ref{cells}b).
In this case, a local dynamical equivalence class similarly consists of four distinct local dimerizations.
This follows from the fact that elementary resonance moves can be carried out around each of the two hexagons contained
in the cell, and moves around different hexagons commute.\cite{misguich}

\subsection{The PEPS parent Hamiltonian}

In Ref. \onlinecite{schuch12}, parent Hamiltonians have been constructed 
utilizing the notion that the state \eqref{RVB} can be written as a PEPS state.
We focus here on the case where the Hamiltonian is the sum over local
projection operators acting on the 19-cite cell discussed above.
Here we first show that the PEPS Hamiltonian constructed for this cell
is identical to the Hamiltonian $H_{19}$ introduced earlier. This is necessary
because we have only shown thus far that the subspace ${\sf RL}({\cal C}_{19})$
projected on by the local terms in $H_{19}$ {\em contains} the support of the local
density matrix, but we have not shown identity. Hence one might suspect that PEPS-Hamiltonian
could be different from $H_{19}$, in that its local operators project onto a smaller, more optimized subspace,
but we will show here that this is not the case.
We do this by first describing the ground state subspace of local terms in the PEPS parent Hamiltonian,
and show it to be identical to ${\sf RL}({\cal C}_{19})$.

In the PEPS formalism, the parent Hamiltonian is constructed from local projections
onto a subspace that is the range of a map from a virtual space into the
local Hilbert space of ``real'' degrees of freedom. This map is described by
a tensor associated to the local cell. The details are given in Refs.\onlinecite{schuch10, schuch12}.
Here we restrict ourselves to the necessary definitions that determine the subspace in question.
We thus assign to each site of the kagome lattice two 3-qutrit states, one associated to each adjacent triangle.
For each triangle, we may define the state
\begin{equation}\label{epsilon}
   \ket{\epsilon}= \sum_{i,j,k=0}^2\epsilon_{ijk}\ket{ijk}+\ket{222}
\end{equation}
Consider now a 19-site double star cell, where for each boundary site, both qutrits are included.
For this cell we now consider virtual states of the following form:
\begin{equation}
    \ket{\phi_{i_1\dotsc i_8}} =\ket{\epsilon}^{\otimes 10} \otimes \ket{i_1\dotsc i_8},
\end{equation}
i.e., a state where each of the ten triangles of the cell is in the state
$\ket{\epsilon}$, and {\em each} of the eight boundary qutrits that are not part
of a triangle of the cell is assigned a qutrit value $i_j \in \{0,1,2\}$.
For each site we now introduce an operator $P$ that maps the two
adjacent qutrits onto a real spin $1/2$ degree of freedom associated with this site:
\begin{equation}\label{P}
    P=\ket{\downarrow} (\bra{02}+\bra{20})+\ket{\uparrow} (\bra{12}+\bra{21}).
\end{equation}
 In essence, therefore, a virtual $\ket{0}$ signifies a down-spin at the same site,
a virtual $\ket{1}$ signifies an up-spin, whereas a virtual $\ket{2}$ signifies the absence
of a valence bond belonging to the triangle associated to the qutrit and touching the site
at which the qutrit resides.
Now we consider the operator $P^{19}$, the tensor product of 19 copies of the operator
$P$ applied to the 19 sites of the cell. We will only be interested in the restriction of this operator
to the virtual subspace spanned by the states $ \ket{\phi_{i_1\dotsc i_8}}$, which is parameterized
by the eight virtual boundary qutrits, and is thus isomorphic to $(\mathbb{C}^3)^{\otimes 8}$. 
The subspace onto which local terms in the PEPS parent Hamiltonian project 
is just the range of the operator $P^{19}$, i.e., the local subspace spanned by
the state $P^{19} \ket{\phi_{i_1\dotsc i_8}}$. Therefore, to show that this Hamiltonian
is identical to $H_{19}$ as defined above through local RVB states,
all we need to demonstrate is that there exists, essentially, a one-to-one correspondence
between the states $P^{19} \ket{\phi_{i_1\dotsc i_8}}$ and the local RVB states
$\psi_{[D],f}$, for the 19-site cell defined earlier. We will do so by observing that the
$0$'s and $1$'s among the entries $i_1\dotsc i_8$ define a free spin configuration $f$,
and that the $2$'s define a class $[D]$ of local dimer patterns connected by resonance moves.
The states $\psi_{[D],f}$ and $P^{19} \ket{\phi_{i_1\dotsc i_8}}$ thus identified are indeed the same.

Before we give some details, we
must make this one-to-one correspondence more precise.
We observe that for  $P^{19} \ket{\phi_{i_1\dotsc i_8}}$ to be non-zero,
the indices ${i_1\dotsc i_8}$ must contain an odd number of $2$'s.
We see this by noting that any qutrit configuration surviving the action of $P^{19}$ must have
exactly one qutrit valued $2$ at each site, i.e., nineteen $2$'s altogether. The state $\ket{\phi_{i_1\dotsc i_8}}$
has the property that on the {\em inside} of each of the ten triangles, there is always an odd number of $2$'s,
which guarantees that the total number of $2$'s {\em inside} those triangles is even.

\begin{figure}[t]
\centering
\includegraphics[width=8cm]{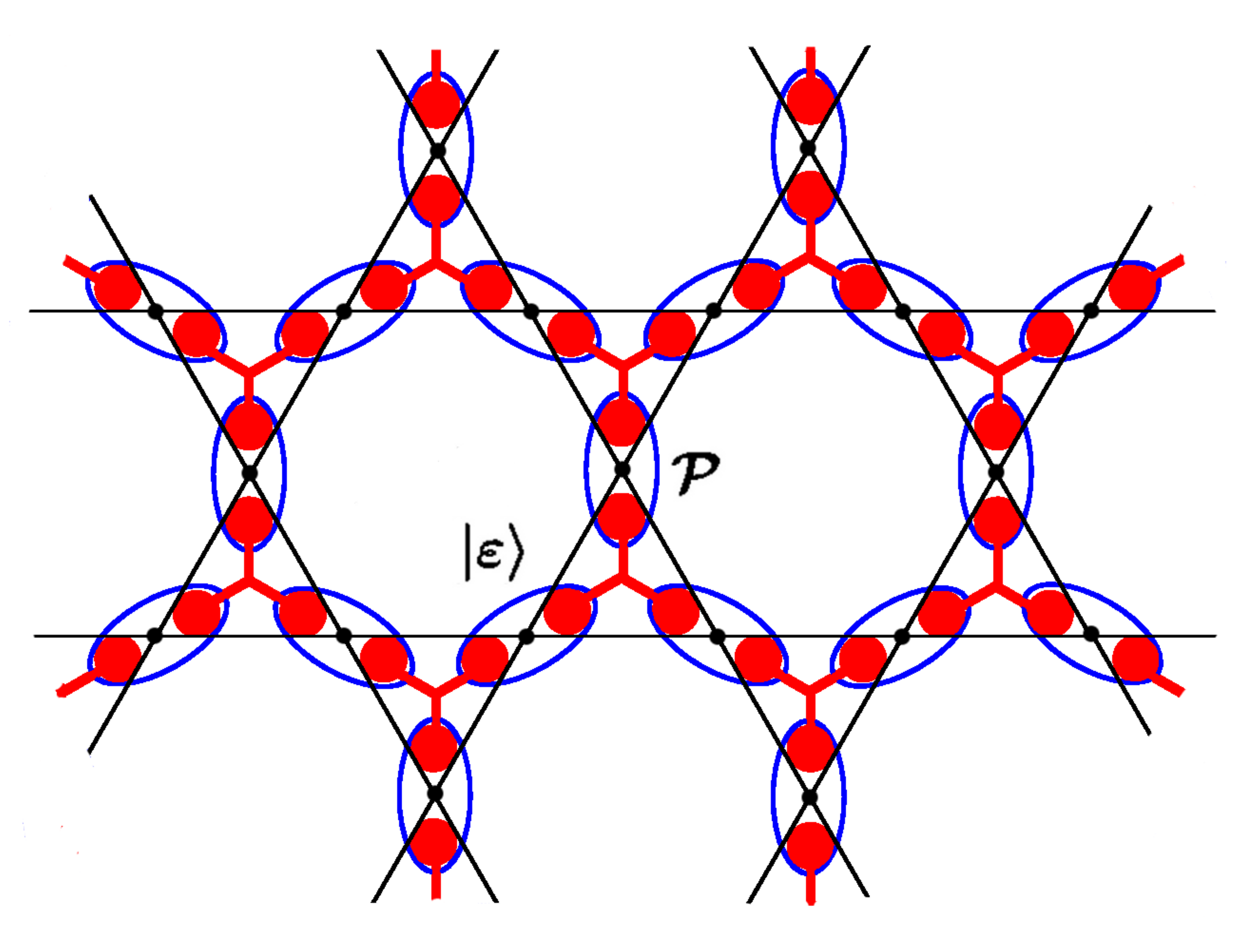}
\caption{
\label{pic4}
[Color online]
The tensor network underlying the PEPS representation of the RVB state, following Ref.
\onlinecite{schuch12}. Red tri-stars denote the state $\ket{\epsilon}$, \Eq{epsilon}, blue ovals the 
projection operator $P$, \Eq{P}.
}
\end{figure}

We must thus have an odd number of $2$'s sitting outside of those triangles, which is by definition
the number of $2$'s among the indices $i_1\dotsc i_8$.
We may thus further restrict the operator $P^{19}$ to the linear span of the states
$\ket{\phi_{i_1\dotsc i_8}}$ with $Z\ket{\phi_{i_1\dotsc i_8}}=-\ket{\phi_{i_1\dotsc i_8}}$,
where $Z$ is the generator of $\mathbb{Z}_2$ represented as $(-1)^{n_2}$, and $n_2$ is the
operator that counts the number of $2$'s in the virtual state.
The one-to-one correspondence, via the map $P^{19}$, which we seek to establish is thus one between
the states $\psi_{[D],f}$ and {\em those} states $\ket{\phi_{i_1\dotsc i_8}}$
whose quantum number $Z$ equals $-1$.
   
To complete the argument, it is now clear that any configuration of virtual boundary 
 qutrits   ${i_1\dotsc i_8}$ with $Z=-1$ specifies a possible configuration
 of free boundary spins $f$ in a state $\psi_{[D],f}$, via the identification $0\equiv\downarrow$,
 $1\equiv\uparrow$. The virtual $2$'s among the qutrits  ${i_1\dotsc i_8}$, on the other hand,
 specify those boundary sites that participate in resonating valence bonds.
 It is easy to see that specifying the boundary sites that participate in valence bonds
 precisely identifies a class $[D]$ of local valence
 bond configurations.\footnote{We could have used this as the definition for a class $[D]$.
 However, in keeping with the definition of Ref. \onlinecite{seidel}, we have defined
 $[D]$ through local ``resonance moves''. It is easy to see that the two definitions coincide.
 For, consider any two local valence bond configurations on the 19-site cell that touch upon
 the same sites. Then their overlap graph must satisfy one of the following three conditions: i)
 there are only short (length 2) loops, ii) there is one long loop encircling exactly one of the two hexagons,
 iii) there is one long loop encircling both hexagons. These three cases correspond, respectively,
 to situations where the dimer configurations are i) identical, ii) related by a single resonance move, iii)
 related by two (commuting) resonance moves. I.e., they are in the same class $[D]$ as defined by resonance moves.
 Easy extension of this argument shows that resonance moves on the kagome are ergodic within topological sectors,
 as first stated in Ref. \onlinecite{misguich}.}
 With $[D]$ and $f$ thus given by the virtual labels  ${i_1\dotsc i_8}$, we indeed have
 (with appropriate normalization conventions) 
 \begin{equation}
 P^{19}   \ket{\phi_{i_1\dotsc i_8}}=\psi_{[D],f}
 \end{equation}
 This establishes the desired identity between the PEPS parent-Hamiltonian for the 19-cite cell
 and $H_{19}$.
   
We note in passing that, for the theory of Ref. \onlinecite{schuch12}, it was of some importance that the map $P^{19}$   
   has a left inverse when restricted to the subspace spanned by the states $\ket{\phi_{i_1\dotsc i_8}}$ with $Z=-1$.
   This has been coined ``$\mathbb{Z}_2$-injectivity''.\cite{schuch10,schuch12} We see here that this is equivalent to the
   linear independence of the states $\psi_{[D],f}$, which in turn is a consequence of the linear independence of the
   states $\ket{D}\otimes \ket{f}$ in \Eq{psiDf} observed in Ref. \onlinecite{seidel}. There, this latter property had been
   used to prove the linear independence of global nearest neighbor valence bond states on general kagome-type lattices.
    
\section{Ground state uniqueness for the parent Hamiltonian $H_{12}$ \label{results}}
   
For definiteness, in the following we will always use the definition of parent Hamiltonians 
given  in Ref. \onlinecite{seidel}, as reviewed in Sec. \ref{locRVB}. For the
19-site cell, this is equivalent to the PEPS definition, as shown above.
In Ref. \onlinecite{schuch12},  it has been shown that for $H_{19}$ and a kagome lattice
with periodic boundary conditions, the ground state space
is spanned by the four topologically degenerate RVB wave functions \eqref{RVBi},
i.e., the four RVB states are the unique ground states of $H_{19}$, up to linear combinations.
The frustration free character of the Hamiltonians $H_{19}$ and $H_{12}$ now allows 
for a simple criterion that is sufficient in order for $H_{12}$ to ``inherit'' this ground state uniqueness-property from $H_{19}$. In PEPS-terminology, we demonstrate an ``intersection property'' \cite{schuch10} for the ground state spaces of $H_{12}$ and $H_{19}$.

Consider a lattice consisting just of one 19-site cell ${\cal C}_{19}$. On this lattice, $H_{19}$ is equal to (the negative of)
 just a single 
projection operator onto the space $ {\sf RL}({\cal C}_{19})$, $H_{19}=-P_{{\cal C}_{19}}$, whereas $H_{12}=H_L+H_R$, with
$H_L$, $H_R$ each being minus the projection operator onto the space
$ {\sf RL}({\cal C}_{12})$, $-P_{\cC_{12}}$, for the left/right 12-site star of the cell, respectively.
The claim is now that if for this 19-site cell, every ground state of $H_{12}$
is also one of $H_{19}$, then this is also true on for any larger kagome lattice
that can be covered by 19-site cells.
For, if $\ket{\psi}$ is the ground state of $H_{12}=-\sum_{\cC_{12}} P_{\cC_{12}}$, 
for some such lattice, then $\ket{\psi}$ is a ground state of each individual operator
$-P_{\cC_{12}}$ in the sum, and therefore also of each operators $H_L+H_R$ defined
as above for {\em any particular} 19-cite cell $\cC_{19}$ of the lattice.
Then, assuming that we can show that any ground state of $H_L+H_R$
is also a ground state of $-P_{\cC_{19}}$, the state $\ket{\psi}$ 
must also be a ground state of $H_{19}$, since the last argument 
can be made for any 19-site cell of the lattice.
Hence $H_{12}$ cannot have more ground states than $H_{19}$,
and, by construction, has the same four RVB ground states \eqref{RVBi}.
This only relies on the statement that any ground state of $H_L+H_R$
is also a ground state of $-P_{\cC_{19}}$, which is apparently a local statement,
i.e., it can be checked on a 19-site lattice.

For the 19-site cell, the set of states $\psi_{[D],f}$ spanning the space ${\sf RL}({\cal C}_{19})$
consists of \RVB19 states. \footnote{This number is easily obtained as $(8 \choose 1)*2^1 + (8 \choose 3)*2^3 + (8 \choose 5)*2^5 + (8 \choose 7)*2^7$, e.g., by counting the number of allowed virtual indices in the states $\ket{\phi_{i_1\dotsc i_8}}$.}They are linearly independent.\cite{seidel}
This is therefore the ground state degeneracy of  $-P_{\cC_{19}}$.
As is elementary to see, each of the states $\psi_{[D],f}$ (for the 19-site cell)
is also a ground state of $H_L+H_R$. The ground state degeneracy
of  $H_L+H_R$ can therefore be only greater than or equal to that of $-P_{\cC_{19}}$,
with the equality implying identical ground state spaces.
We have shown that this is indeed that case, using two different numerical methods.
The first is by straightforward diagonalization, 
using both $S_z$-conservation and the two mirror symmetries
of the 19-site cell. The second is specific to finding the ground state subspace
of frustration free Hamiltonians.
Since any ground state of $H_L+H_R$ must be a ground state of both
$H_L$ and $H_R$, we may first obtain a complete set of ground states
of $H_L$, working only on the 12-site cell. These are just the states
$\psi_{[D],f}$ for the 12-site cell. (These are likewise linearly independent,\cite{seidel}
which leads to a $\mathbb{Z}_2$-injectivity property for the 12-site cell noted
in Ref. \onlinecite{schuch12}, which is analogous to that for the 19-site cell already mentioned.)
Let us denote the latter by $\psi_{[D],f}^{12}$ to emphasize that these are states of 12-spins.
On the 19-site cell, the ground states of $H_L$ are thus of the form $\ket{\psi_{[D],f}^{12}}\otimes\ket{r}$,
where $\ket{r}$ is an arbitrary configuration of the remaining spins of the 19-site double star cell
that do not belong to the left star. The number of these states is now rather small
compared to the full dimension $2^{19}$ of the Hilbert space of the 19-cite cell.
The ground states of $H_L+H_R$, being ground states of {\em both} $H_L$ and $H_R$,
must now be a linear combination of the states $\ket{\psi_{[D],f}^{12}}\otimes\ket{r}$.
We may thus diagonalize  $H_R$ within this subspace. By the variational principle,
eigenstates within this subspace that have the lowest possible eigenenergy of $-1$ 
are true eigenstates also of the unrestricted $H_R$, and correspond to ground states 
of $H_L+H_R$. Conversely, every ground state of $H_L+H_R$ can be obtained in this
two step procedure, which allows one to avoid ever working with the full Hilbert space
of dimension $2^{19}$. ($S_z$-conservation may further be used in this case as well.)
We have used both of the above methods to confirm that the ground state degeneracy
of $H_L+H_R$ is identical to that of $-P_{\cC_{19}}$, thus the ground state spaces of these
two operators are identical.

The above observations complete the demonstration that on any finite kagome lattice
that can be covered by 19-cite cells, the ground state spaces of $H_{12}$ and $H_{19}$
are identical. Hence for any such lattice to which the proof of Ref. \onlinecite{schuch12}
applies (in particular for the periodic boundary condition chosen there),
the ground state space of $H_{12}$ is spanned by the fourfold topologically degenerate
RVB-states \eqref{RVBi}. This property of  $H_{12}$ had
  originally been conjectured in Ref. \onlinecite{seidel}, where it was proven to hold
only within the subspace of nearest neighbor valence bond coverings.
We may ask if the method described here can be applied to different ground states
and their parent Hamiltonians. A natural modification of the RVB states is to utilize a different 
sign convention. Hence, replace \Eq{RVBi} with
\begin{equation}\label{TRVBi}
   \widetilde{\ket{{\sf RVB},i}}= \sum_{D\in[i]}  (-1)^{\hat N}  \ket{D}\;.
\end{equation}

Here, $\hat N$ is an operator that counts the number of resonance moves required to
transform the dimer covering $D$ into a given reference dimer covering in the topological sector $[i]$.
That this is well defined (modulo 2) can be seen from the following, alternative definition.
Pair up all hexagons on the lattice, say, into nearest neighbor pairs.
Then connect members of a pair through paths that start and end at the midpoints of the respective
hexagons, and intersect links of the kagome lattice, avoiding sites.
The operator $\hat N$ may then be defined as the number of dimers crossed by these paths
 in the dimer covering $D$. (By the linear independence of the states $\ket{D}$,\cite{seidel}
 this is indeed a well defined operator within the subspace of nearest neighbor valence bond states, 
 although this is perhaps not essential in defining the state
 \eqref{TRVBi}, where it is sufficient that a phase $(-1)^{N_D}$ can be associated to each dimer covering $D$.)
 It is easy to see that $(-1)^{\hat N}$ changes sign upon a dimer resonance move around any hexagon. 
 This construction can be thought of as a variational excited state for the original parent Hamiltonian (having the states \eqref{RVBi} as ground states), 
 where a ``vison'' type excitation is placed in each hexagon.
 
 One may instead want to construct a new parent Hamiltonian for this new variational wave function.
 It is easy to generalize the construction of parent Hamiltonians
 given in Sec. \ref{locRVB} to the present situation.
 For the 12-site version of the parent Hamiltonian,
 all one needs to do is to replace the generating set of ${\sf RL}(\cC_{12})$,
 \Eq{RVB12}, with
 \begin{equation}\label{TRVB12}
   \langle \{ (\ket{D} - \ket{D^\ast} ) \otimes \ket{f} \} \rangle\,.
\end{equation}
Similarly, an additional sign could be introduced as described above
in the generating set of the subspace 
  ${\sf RL}(\cC_{19})$. We denote the resulting parent Hamiltonians by
  $\widetilde H_{12}$ and $\widetilde H_{19}$, respectively.
  We have shown using the same methods described above  in this
    section that the ground state spaces of   $\widetilde H_{12}$ and $\widetilde H_{19}$
    are identical. We do not know at present if the methods developed in Ref. \onlinecite{schuch12}
    to show uniqueness of the fourfold degenerate RVB ground states carry over to 
    $\widetilde H_{19}$. We note that one essential ingredient of these methods, the $\mathbb{Z}_2$-injectivity
    of the tensor associated with the 19-cite cell, still holds in this case.
    This is so for reasons identical to those given above for the original RVB-states, namely,
    the linear independence of the states $\ket{D}\otimes\ket{f}$ for the 19-site cell.
    For the 12-site cell, however, the corresponding $\mathbb{Z}_2$-injectivity
    no longer holds. This is so since the generating states of the subspace ${\sf RL}(\cC_{12})$
    listed in \Eq{TRVB12} are no longer linearly independent, as already noted in Ref. \onlinecite{seidel}.
    We leave further investigation of the ground state uniqueness of  $\widetilde H_{12}$ and $\widetilde H_{19}$
    for future work.

\section{conclusion \label{conclusion}}

In this work, we have clarified connections between
parent Hamiltonians for nearest neighbor resonating valence bond
states arising from the PEPS construction and from earlier considerations.
The case that these Hamiltonians 
demonstrate the existence of SU(2) invariant topological spin liquids
rests primarily on the nature of correlations in their ground state, 
as well as the
uniqueness of these ground states, i.e., in particular, the fact that they display the 
correct degeneracy on any finite lattice.
This ground state uniqueness was first proven for a Hamiltonian that acts
on 19 spins a time.\cite{schuch12}
Here we have combined the latter result with numerical work on finite clusters
to establish this ground state uniqueness for a simpler 12-site Hamiltonian,
completing the prove of an earlier conjecture. Technically, this was done
by demonstrating a ground state intersection property for the finite clusters studied.
We believe that this 12-site Hamiltonian is the smallest parent Hamiltonian
for the prototypical nearest neighbor resonating valence bond state \eqref{RVB} on the kagome
lattice that has these desired features. Our results should also be of interest
in the broader context of frustration free two-dimensional Hamiltonians.

\begin{acknowledgements}
This work has been supported by the National Science
Foundation under NSF Grant No. DMR-1206781.
AS would like to thank N. Schuch for insightful discussion and hospitality
during preliminary stages of this work, and moreover, performing independent confirmation
of our main result upon receipt of our final draft.
\end{acknowledgements}

\bibliography{unique}

\end{document}